\documentclass[pre,twocolumn, showpacs]{revtex4}
\usepackage{graphicx}
\usepackage{amsmath,amssymb,bm}
\usepackage{times}
\usepackage{psfrag}
\usepackage{verbatim}
\usepackage{color}
\newcommand{\expt}[1]{\langle #1 \rangle}

\begin{document}
\title{Using force covariance to derive effective stochastic interactions in dissipative particle dynamics}
\author{Anders Eriksson}
\author{Martin Nilsson Jacobi}
\author{Johan Nystr\"{o}m}
\author{Kolbj\o rn Tunstr\o m}
\affiliation{Complex Systems Group, Department of  Energy and Environment, Chalmers University of Technology, 41296 G\"oteborg, Sweden} 


\begin{abstract}

There exist methods for determining effective conservative interactions in coarse grained particle based mesoscopic simulations. The resulting models can be used to capture thermal equilibrium behavior, but in the model system we study do not correctly represent transport properties. In this article we suggest the use of force covariance to determine the full functional form of dissipative and stochastic interactions. We show that a combination of the radial distribution function and a force covariance function can be used to determine all interactions in dissipative particle dynamics. Furthermore we use the method to test if the effective interactions in dissipative particle dynamics (DPD) can be adjusted to produce a force covariance consistent with a projection of a microscopic Lennard-Jones simulation. The results indicate that the DPD ansatz may not be consistent with the underlying microscopic dynamics. We discuss how this result relates to theoretical studies reported in the literature.

\end{abstract}

\maketitle

\section{Introduction}

It is the long-standing aim of molecular simulations to elucidate mechanisms that cannot be directly observed in experiments, or understood in terms of more abstract models. Though extremely successful in many areas, when applied to mesoscopic systems, such as membranes or complex fluids, one often finds that the relevant time and length scales on which important mechanisms take place are beyond the reach of direct, detailed, simulations. As a consequence, several coarse-grained methods have been developed to allow for a larger time span to be simulated.
Lattice gases \cite{frisch} corresponds to dividing the system into an array of sub-systems, each a thermodynamic system on its own with a local temperature, pressure, particle density and velocity distribution. Other coarse-graining procedures have explicit particles with pair-wise interactions; well-known examples are united atoms \cite{rapaport97}, smoothed particle hydrodynamics (SPH) \cite{monaghan} and dissipative particle dynamics (DPD) \cite{hoogerbrugge_koelman92}. There also exists hybrid methods suitable for example in molecular simulations where atomistic resolution is needed only in spatially localized domains, e.g. \cite{Praprotnik2005, Praprotnik2007a}. 

Atomic force fields for molecular dynamics (MD), derived from potentials defined empirically or theoretically (e.g. from quantum-mechanical models), are relatively mature. In contrast, it is much less clear how to choose the effective force fields for coarse-grained models, partly because the connection between the degrees of freedom in the coarse-grained dynamics and the underlying molecular dynamics differ from one coarse-graining procedure to the next. 
Frequently, simple heuristic forces are used; partly because of computational ease but also because the detailed forces may not be known \cite{groot_warren97}. The magnitude of the forces are then chosen to match macroscopic observables of the system, such as the compressibility.
\citet{forrest_suter95} (see also references therein) calculate an effective force on the particles, interacting according to the Lennard-Jones potential, which correspond to the average effect of the true forces during a time interval. Because the time-averaged force is effectively an average over rapid fluctuations of close particles, the effective potentials are much softer than the Lennard-Jones potential at small distances.
An alternative approach, which we will use later in this paper, is to use the fact that for particle systems with central forces there is a one-to-one relation between the radial distribution of particles at thermal equilibrium and the pair-wise potential \cite{Henderson}.

When coarse-graining a molecular system, the effective interactions in the resulting system can either be deterministic (smooth, due to averaging of fast degrees of freedom), in which case just matching the equilibrium properties of the system gives the correct dynamics, or the fast degrees of freedom act as a noisy driving force. Which of these two scenarios that best describe the system at hand depends on the exchange of energy between the simulated system and the degrees lost in the coarse-graining procedure. If a substantial amount of energy is exchanged then the motion of the coarse-grained particles will not be smooth or deterministic. In this case, it is not sufficient to capture the conservative forces but we must also introduce dissipative and stochastic forces. These forces are included in the SPH and DPD models. The models are quite similar, and in this article we choose to focus on DPD. As mentioned in the previous paragraph the smooth, or conservative, part of the interaction can be determined from the radial distribution function in in thermal equilibrium \cite{Henderson}. In other words, the radial distribution at equilibrium is not affected by the noise term in the interaction (provided an appropriate deterministic dissipative force is added, ensuring  the equilibrium temperature). Clearly it follows that equilibrium properties that are determined by the radial distribution are also not affected. Examples are compressibility and other observables defining the equation of state. Other characteristics such as the diffusion coefficient and the viscosity do depend on the stochastic interactions. To capture these properties it is central to choose the stochastic interaction, including its radial dependence, as correctly as possible.

To accurately describe the dissipative and stochastic part of the dynamics we must introduce a new observable, complementing the radial distribution function. A candidate could be the autocorrelation of the velocity, which is directly related to the stochastic driving on a single particle as well as the diffusion coefficient through Green-Kubo relations \cite{green54,kubo57}. In a particle-based mesoscopic model such as DPD, the stochastic interaction is represented as a force between the particles. Clearly this interaction must have a radial dependence (typically with short range). A useful observable for estimating the stochastic interaction must therefore be able to resolve the radial dependence. In this paper we suggest the use of the force covariance function as a candidate for such an observable, and use DPD as a test case for the method.

\section{Dissipative particle dynamics}

DPD was introduced in 1992 by \citet{hoogerbrugge_koelman92} as a simulation technique for hydrodynamic phenomena. The method has received much theoretical attention~\cite{espanol_warren95,coveney_espanol,marsh1997_1,marsh1997_2,marsh1997_3} that provides support for this kind of model and has been established as a standard method for mesoscopic simulation. Amongst other things DPD has been used to  study complex fluids \cite{groot_warren97}, spontaneous self-assembly of amphiphilic molecules into bilayered membranes \cite{shillcock_lipowsky02}, vesicles \cite{yamamoto_etal02,yamamoto_hyodo03}, and hydrodynamics \cite{trofimov_etal02}. In its standard form, DPD is a particle model with pairwise interactions, quite similar to molecular dynamics, but with a dissipative and stochastic contribution to the interactions between the particles. 

In its simplest form, the equations of motion for a DPD model, with mesoscopic particles positioned at $\mathbf{r}_{i}$, with velocities  $\mathbf{v}_{i}$ and momenta $\mathbf{p}_{i}$ can be written as a system of Langevin equations
\begin{align}\label{eq:simple_langevin}
	\dot{ \mathbf{r} }_{i}    &= \mathbf{v}_{i}, \nonumber \\
	\dot{ \mathbf{ p } }_{i}  &= \sum_{j \neq i} \left[ \mathbf{F}_{ij}^\text{C} + \mathbf{F}_{ij}^\text{D}
										  +\mathbf{F}_{ij}^\text{S} \right],
\end{align}
where $\mathbf{F}_{ij}^\text{C}$, $\mathbf{F}_{ij}^\text{D}$ and  $\mathbf{F}_{ij}^\text{S}$ are the conservative, dissipative and stochastic forces between particles $i$ and $j$. Both the conservative and non-conservative interactions in DPD are modeled by central forces obeying Newton's third law, ensuring that (angular) momentum is conserved~\cite{hoogerbrugge_koelman92}. The dissipative and stochastic forces are
\begin{align}
	\mathbf{F}_{ij}^\text{D} &= 
	- \omega^{D}(r_{ij}) \, \mathbf{e}_{ij} \!\cdot\!(\mathbf{v}_i - \mathbf{v}_j)\, \mathbf{e}_{ij}, \\
	\mathbf{F}_{ij}^\text{S} &=  \omega^{S}( r_{ij} ) \, \zeta_{ij} \, \mathbf{e}_{ij},
\end{align}
where $r_{ij}$ is the distance between particles $i$ and $j$, and $\mathbf{e}_{ij} $ is the unit vector pointing from $j$ to $i$. The scalar functions $\omega^\text{D}(r_{ij})$ and $\omega^\text{S}(r_{ij})$ describe friction and noise, respectively.  $\zeta_{ij}$ is interpreted as a symmetric Gaussian white noise term with mean zero and covariance
\begin{equation}
	\expt{\zeta_{ij}(t)\zeta_{i'j'}(t')} = (\delta_{ii'}\delta_{jj'} + \delta_{ij'}\delta_{ji'})\delta(t-t'), 
\end{equation}
where $\delta_{ij}$ and $\delta(t)$ are the Kronecker and Dirac delta functions. Assuming that the equilibrium distribution of a DPD system is given by the canonical ensemble, the fluctuation-dissipation theorem leads to a relation between the dissipative and stochastic parts \cite{espanol_warren95}:
\begin{equation}\label{eq:fluctdisstheorem}
	\omega^\text{D}(r) = (2 k_{B}T)^{-1}[\omega^\text{S}(r)]^{2}.
\end{equation}
For simplicity, we drop the superscript and write $\omega(r) \equiv \omega^\text{S}(r)$. Eqs.~(\ref{eq:simple_langevin})--(\ref{eq:fluctdisstheorem}) together establishes the general form of the DPD dynamics. Both the conservative force $\mathbf{F}_{ij}^\text{C}$, or equivalently the corresponding scalar potential, and the scalar function $\omega(r)$ depend on the particular system of interest and need to be determined to obtain the correct DPD model. In practice, this is the difficult part of DPD, and also the rationale behind the heuristic approach in deciding the interactions. As an example, the common practice for fluid like systems is to apply linear functions with a cutoff radius $r_{c}$;
\begin{align}
	\mathbf{ F }_{ij}^\text{C} &= 	(1 - r_{ij}/r_c) a_{ij} \chi_{ij}  \, \mathbf{e}_{ij} = F^\text{C}(r_{ij})  \, \mathbf{e}_{ij}, \label{eq:linear_force}\\		
	\omega(r_{ij})                &= (1 - r_{ij}/r_c) \sigma \chi_{ij} ,	\label{eq:linear_noise}				\end{align}
where $a_{ij}$ is the strength of the conservative force between particles $i$ and $j$, $\sigma$ is the amplitude of the noise, and $\chi_{ij}$ is one for $r_{ij}\leq r_{c}$ and zero elsewhere. 

Despite its popularity and theoretical support, it is unclear how DPD should be interpreted as a coarse-grained model~\cite{martys_2004}. One point of view, and the one we will elaborate on in this article, is to consider DPD as a systematic coarse-graining of an underlying atomistic system. If the DPD method could be shown to have a firm microscopic foundation, that would greatly impact our ability to couple DPD to actual physical systems. Several authors, e.g.~\cite{flekkoy1,flekkoy2,flekkoy3,espanol_review}, have established bottom-up connections between the micro- and mesoscale and obtained mesoscopic dynamics resembling DPD. The resulting methods differ from DPD as they incorporate the geometry of the system in the equations, implying forces that are not central or pairwise, while DPD is a model with only pairwise interactions. It is clear that the validity of DPD as a coarse-grained model, or how well DPD represents an underlying microscopic system, has not been fully resolved. 

To obtain a well-defined bottom-up scheme, the dynamics of the coarse-grained DPD particles must be defined through a projection of the microscopic trajectories. The problem is to find a closed representation of the system at the coarse-grained level, i.e. to determine all interactions in the DPD model. In this article we investigate a method of estimating the DPD interactions using measurements on the coarse-grained level of a simulation. By applying the method to a typically assumed projection of a microscopic system, we clarify some important aspects of DPD as a systematically coarse-grained model. 

The DPD technique has its theoretical foundations in Mori-Zwanzig theory on projection operators~\cite{zwanzig,zwanzig_book,mori1,mori2}. In short, the theory states that given a microscale dynamics, a lower dimensional representation can be  formally attained through a projection of the phase space, where fast degrees of freedom are treated as  Markovian (white) noise~\cite{zwanzig}. This framework can be applied to molecular dynamics~\cite{Junghans, Soddemann}. Naturally, how faithfully the coarse-grained model will represent the underlying dynamics depends on the choice of projection. The DPD method assumes a projection resulting in a mesoscopic model characterized as a particle based Langevin dynamics with pairwise and negated central forces. The internal degrees of freedom in the mesoscopic particles give rise to dissipation and noise, which is captured by non-conservative pairwise interactions. As a consequence, sufficiently close to equilibrium, one obtains the classical result of the asymptotic $t^{-d/2}$ decay of the velocity auto-correlation ($d$ is the dimensionality of the system) \cite{ernst_etal71}. In addition, the interactions give rise to hydrodynamic modes in the fluid \cite{zwanzig_bixon70,ernst_etal71}, which lead to the Navier-Stokes equations on the macroscopic level \cite{espanol_hydrodynamics}. 

\section{Estimating the effective forces}

Given the DPD ansatz for the effective equations of motion, the question is: how does one find the conservative and dissipative forces $F^\text{C}(r)$ and $\omega(r)$? In this section we present the theoretical motivations for our method, and apply it to DPD simulations to test the accuracy of the method on a case where we know the ansatz to be true. In section~\ref{sec:lj_example} we apply the method to a coarse-graining of a system of particles interacting via the Lennard-Jones potential, in order to see how the method fares on a classical molecular dynamics system.

\subsection{The conservative force term}

The original motivation \cite{groot_warren97} for a repulsive conservative force was a measurement of the effective potential for the interaction between particles in a Lennard-Jones fluid \cite{forrest_suter95}. More generally applicable methods for estimating the conservative interactions are based on the radial distribution function (RDF) in thermal equilibrium \cite{Lyubartsev2003, Ilpo, Soper, Reith, Izvekov, Almarza_Lomba_PRE}. In these reports, the estimate of the conservative force is obtained using a result by \citet{Henderson}, stating that the difference between two pairwise potentials that give rise to the same radial distribution function must be a constant gauge shift, and hence of no physical significance. The importance of this theorem lies in the one-to-one correspondence between potential and radial distribution function.

The conservative interactions are determined by the RDF alone, which in turn is determined by the thermal equilibrium of the system. As long as the fluctuation-dissipation theorem holds, the thermal equilibrium is independent of the specific form of the dissipative and random interactions \cite{espanol_warren95}, and it follows that we can estimate the conservative forces from a given RDF independently of the stochastic forces. Here we use the inverse Monte Carlo method of Lyubartsev and Laaksonen~\cite{Lyubartsev1995}, which starts from a Boltzmann ansatz of the potential and then, through iteration, finds a potential giving rise to the desired RDF. In what follows we briefly describe the method, following~\cite{Lyubartsev1995}.
 
The connection between the RDF and the potential can be found from the Hamiltonian of the system. Consider a system of particles with pairwise interactions. It can be discretized as
\begin{equation}
 H = \sum_{\alpha} \Phi_{\alpha}S_{\alpha},
\end{equation}
which corresponds to using a stepwise constant potential, $\Phi_\alpha$. $S_\alpha$ denotes the number of particle pairs separated by a distance in the range $r_\alpha$ to $r_{\alpha+1}$, where $r_0 = 0$, $r_1 = dr$ and $r_\alpha = \alpha \cdot dr$. The average of $S_\alpha$ is directly connected to the radial distribution function, $g(r)$, by the relation
\begin{equation}\label{eq:S-g relation}
	\frac{\expt{S_\alpha}}{N(N-1)/2} = \frac{V_\alpha}{L^3} g(r),
\end{equation}
where $N$ is the number of particles, $L^3$ the volume of the simulation box and $V_\alpha$ the volume of the spherical shell between radii $r_\alpha$ and $r_{\alpha+1}$. Using a Monte Carlo (MC) approach, the system may be simulated with a start potential $\Phi_\alpha^{(0)}$. It is common practice to choose this to be the potential of mean force,
\begin{equation}\label{eq:PMF}
	\Phi_\alpha^{(0)} = -k_BT \ln g(r_\alpha). 
\end{equation}
The correspondance between potential and RDF is an equilibrium result and hence only valid for fixed temperatures and densities. These quantities must therefore be the same in the MC simulation as they were in the original simulation from which the RDF was obtained.

Simulating with the trial potential, $\Phi_\alpha^{(0)}$, produces an $\expt{S_\alpha^{(0)}}$ which may differ from the correct value, $S_\alpha^*$. The difference, $\Delta\expt{S_\alpha}^{(0)} = \expt{S_\alpha}^{(0)} - S_\alpha^*$, is used to find a new trial potential by solving for $ \Delta \Phi$ in the linear equation system
\begin{equation} \label{eq:S-V equation system}
	\Delta\expt{S_\alpha} = \displaystyle \sum_\gamma\frac{\partial \expt{S_\alpha}}{\partial \Phi_\gamma} \Delta \Phi_\gamma,
\end{equation}
with $\frac{\partial \expt{S_\alpha}}{\partial \Phi_\gamma}$ given by~\cite{Lyubartsev1995}
\begin{equation}\label{eq:dsdv}
	\frac{\partial \expt{S_\alpha}}{\partial \Phi_\gamma} = -\frac{\expt{S_\alpha S_\gamma} - \expt{S_\alpha}\expt{S_\gamma}}{k_BT}.
\end{equation}
The next guess for potential is then $\Phi^{(1)} = \Phi^{(0)} - \Delta \Phi$. This procedure is repeated until $\Phi$ has converged to a potential that reproduces the original RDF. 

The potential of mean force is usually a good first approximation to the final potential, and convergence to a unique potential normally takes less than ten updates in the Monte Carlo simulations. This is especially true for the soft coarse-grained potentials we get from considering effective interactions between clusters of particles. For instances that nevertheless require special care, problems with convergence for the potential over successive MC simulations can generally be overcome by moving only a fraction in the direction specified by Eq. (\ref{eq:S-V equation system}). 

\subsection{The dissipative force term} 
\label{sec:estimate_dissipative_term}

Assuming that the DPD ansatz is valid, the functional form of the dissipative term (and through the fluctuation dissipation theorem, Eq. (\ref{eq:fluctdisstheorem}), the stochastic term) can be isolated through a Kramer-Moyal expansion \cite{gardiner04} of Eq.~(\ref{eq:simple_langevin}): 
\begin{align}\label{eq:dpidpj_dt}
	& - \expt{\delta \mathbf{p}_{i} \!\cdot\! \delta \mathbf{p}_{j}}/\delta t
	= \expt{\omega^{2}( r_{ij} )} - 
	      \sum_{k \neq i, l \neq j} \expt{ \mathbf{F}_{ik}^\text{C} \!\cdot\! \mathbf{F}_{jl}^\text{C} }\, \delta t  \nonumber\\
	& -  \sum_{k \neq i, l \neq j} 
	\frac{ \expt{ \omega^{2} ( r_{ik} ) \, \omega^{2} ( r_{jl}) \, (\mathbf{e}_{ik} \!\cdot\! \mathbf{v}_{ik}) \, 
	(\mathbf{e}_{jl} \!\cdot\! \mathbf{v}_{jl}) \, (\mathbf{e}_{ik} \!\cdot\! \mathbf{e}_{jl}) } }{(2k_{B}T)^{2}}  \, \delta t \nonumber\\
	& + \sum_{k \neq i, l \neq j} \frac{ \expt{ \omega^{2} ( r_{ik}) \, (\mathbf{e}_{ik} \!\cdot\! \mathbf{v}_{ik}) \, (\mathbf{e}_{ik} \!\cdot\! \mathbf{F}_{jl}^\text{C}) }}{2k_{B}T} \, \delta t	 \nonumber\\
	& + \sum_{k \neq i, l \neq j} \frac{ \expt{ \omega^{2} ( r_{jl}) \, (\mathbf{e}_{jl} \!\cdot\! \mathbf{v}_{jl}) \, (\mathbf{e}_{jl} \!\cdot\! \mathbf{F}_{ik}^\text{C}) }}{2k_{B}T} \, \delta t	 + O(\delta t^2),	
\end{align}
where $\delta\mathbf{p}_i(t) = \mathbf{p}_i(t+\delta t)-\mathbf{p}_i(t)$. All averages in Eq. (\ref{eq:dpidpj_dt}) are conditioned on that the distance $r_{ij}$ between particles $i$ and $j$ equals $r$. This equation provides a relationship between the functional form of the stochastic and dissipative interactions, $\omega(r)$, and the force covariance, $\kappa_{\text F}$, defined as: 
\begin{equation}
	\kappa_{\text F} \equiv -\expt{\delta \mathbf{p}_{i} \!\cdot\! \delta \mathbf{p}_{j}}/\delta t.
\end{equation}

In the DPD simulations we can take $\delta t$ small enough that only the leading term of Eq.~(\ref{eq:dpidpj_dt}) is significant. If the microscopic dynamics is deterministic, like in most molecular dynamics, there generally exists a time-scale below which the forces are smooth functions of time (this is the time-scale on which the molecular dynamics can be integrated). On this time-scale, the projected dynamics is also smooth (if the projection is smooth) but not autonomous. It follows that for small $\delta t$ the force covariance $\kappa_{\text F}$ is proportional to $\delta t$, corresponding to $\omega(r) = 0$ in Eq.~(\ref{eq:dpidpj_dt}).

In fluids, the magnitude of $\kappa_{\text F}$ will typically increase with increasing $\delta t$, because on relatively short time scales the particles of the fluids oscillate in a cage formed by their closest neighbours (the Franck-Rabinowitch effect). We consider these rapid fluctuations to correspond to fast degrees of freedom in the system. To get an idea of the time-scales involved, consider water particles in a fluid at normal pressure and room temperature. The particles' distance to their closest neighbours oscillate around the first peak in the radial distribution function, at approximately $0.28$~nm. The half-width of the peak, approximately $0.05$~nm, gives an indication of how far the molecule travels before experiencing strong repulsive forces from other particles. We estimate the typical velocity as the root mean square (RMS) velocity
\begin{align}
	v_\text{RMS} = \sqrt{\expt{\mathbf{v}^2}} = \sqrt{3k_\text{B}T/m}.
\end{align}
At room temperature ($25\,^\circ$C), the RMS velocity is approximately $640$~m/s. One may argue that the orientation of the particle velocities are essentially random, so that the RMS difference in velocity is $v_\text{RMS}\sqrt{2}$, and that they collide at half the half-width. The time to travel this distance at the typical velocity is then approximately $0.03$~ps, and we take this as a rough approximation to the timescale at which the fast dynamics occur. It is only at timescales significantly larger than this timescale that we can expect to approximate the fast degrees of freedom with a spatially structured but Markovian noise as in the DPD ansatz. On this time scale, the fluid approaches a local thermal equilibrium on the length scale of the coarse-grained particles, determined by the local concentration, local average velocity and kinetic energy \cite{ernst_etal71}.

As a consequence, it is generally not possible to take the limit of $\delta t \rightarrow 0$ in the numerically estimated $\kappa_{\text F}$ to find $\omega(r)$. Rather, we will assume that there exists a time interval where the fast degrees of freedom can be approximated by noise, and where $\kappa_{\text F}$ is an approximately linear function of time. Given two values of $\delta t$ in this interval, $\delta t_1$ and $\delta t_2$, we can use Richardson extrapolation to eliminate the $\delta t$ term in Eq.~(\ref{eq:dpidpj_dt}) to obtain an $O(\delta t^2)$ estimate for $\omega^2(r)$: 
\begin{equation}\label{eq:omega_richardson}
	\omega^2( r_{ij} )|_{\delta t_0} \approx
	\frac{\delta t_1}{\delta t_2}
	\frac{ \expt{\delta \mathbf{p}_{i} \!\cdot\! \delta \mathbf{p}_{j} } |_{ \delta t_2} }{ \delta t_2 - \delta t_1 } - 
	\frac{\delta t_2}{\delta t_1} 
	\frac{ \expt{\delta \mathbf{p}_{i} \!\cdot\! \delta \mathbf{p}_{j} } |_{ \delta t_1} }{ \delta t_2 - \delta t_1 }.
\end{equation}
An alternative approach is to do a linear fit with respect to $\delta t$ in this region, for each value of $r$, and from the best fit take the intersection with the line $\delta t = 0$.

\subsection{Recreating the effective interactions of DPD simulations} 

At this point we have established the principles behind our method. An important consistency check is to apply the method to standard DPD simulations, where the dynamics is truly Langevinian. This was done by performing DPD simulations with different functional forms of both ${\bf F}_{ij}^\text{C}(r)$ and $\omega(r)$. Using standard DPD units, the simulation region was a periodic cubic box with side length $L = 8.7359$, with $3$ particles per volume unit, giving a total of $2000$ particles. From the simulations, the RDF and $\kappa_{\text F}$ were calculated for $100$ $r$-values in the range $0$ to $1.75$, after which the RDF had converged to $1.0$. The time-step size used in the simulations was small ($\delta t = 10^{-3}$) compared to a normal DPD simulation. The reason for this was to approach the limit of small $\delta t$, so that the terms proportional to $\delta t$ could be ignored in Eq.~(\ref{eq:dpidpj_dt}); $\omega^{2}(r)$ is then given simply by $\kappa_{\text F}$. 

\begin{figure} 
\centering
\includegraphics[width=8cm]{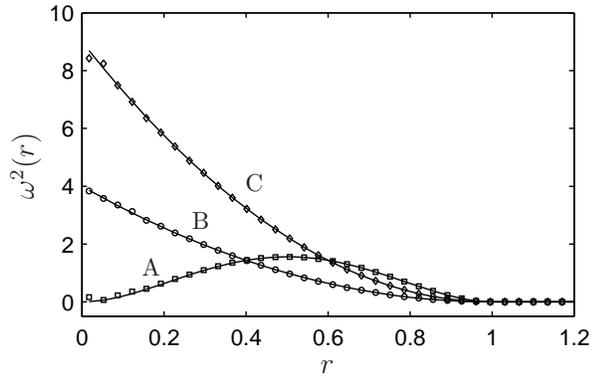}
\caption{\label{fig:dissForce_estimation}
The plots show three different functional forms of $\omega^2(r)$. The symbols ($\diamond$, $\circ$ and ${\scriptscriptstyle \square}$) show the values found by our method: measuring $\kappa_{\text F}$ of a DPD simulation. The exact functional forms used in the simulations are plotted as solid lines. All units are standard DPD units. The conservative and dissipative forces are (for $r \in [0,1]$): 
(A) $\text{F}^\text{C}(r) = 10  (1-r)$, $\omega(r) = 5r(1-r)$
(B) $\text{F}^\text{C}(r) = 10  (1-r)$, $\omega(r) = 2 (1-r)$
(C) $\text{F}^\text{C}(r) = 10 r(1-r)$, $\omega(r) = 3 (1-r)$.   
For $r>1$, both functional forms are zero.}
\end{figure}

\begin{figure}[htp]
\centering
\includegraphics[width=8.0cm,angle=0]{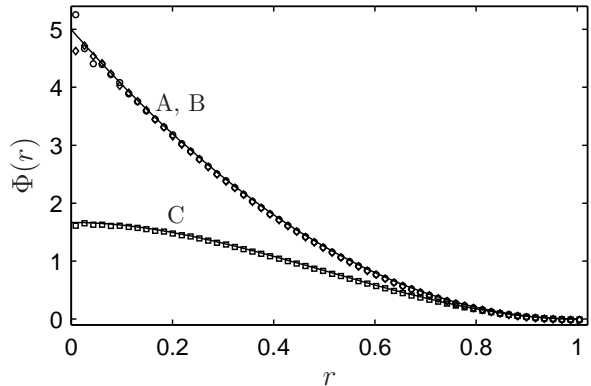}
\caption{The conservative potentials $\Phi(r) \equiv \int_r^\infty \text{d}r' F^\text{C}(r')$ from three DPD simulations have been recreated from RDF-data. Cases (A) and (B) correspond to a linear DPD-force (i.e. quadratic potential) with different random force parts. For case (C), a quadratic conservative force was used. For details, see caption of figure~\ref{fig:dissForce_estimation}. In all three cases, the potential was exactly recreated up to statistical accuracy.}
\label{fig:ThreeCasesDPDpotential}
\end{figure}

In all cases the method accurately recreated the DPD interactions used in the simulations. Fig.  \ref{fig:dissForce_estimation} shows the results of recreating  $\omega^2(r)$ for three different functional forms. The conservative potential was also varied (see figure caption for details), and plots of recreated potentials from these simulations are shown in figure~\ref{fig:ThreeCasesDPDpotential}. 

\begin{figure} 
\centering
\includegraphics[width=8cm]{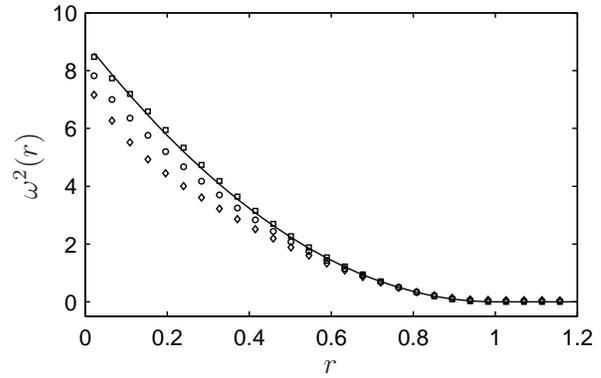}
\caption{\label{fig:dissForce_richardson}
The plot shows two measurements of $\kappa_{\text F}$ obtained from a DPD simulation using time steps of different size ($\circ$: $\delta t_1=0.025$ and $\diamond$: $\delta t_2=0.05$ ). The resulting estimate of $\omega^2(r)$,  obtained by Richardson extrapolation of $\kappa_{\text F}$ measurements, is shown as ${\scriptscriptstyle \square}$. The solid line shows the exact form of $\omega^2(r) = (3(1-r))^2$ used in the simulation. The conservative force used was $\text{F}^\text{C}(r) = 10 r(1-r)$.}
\end{figure}

\begin{figure}
\centering               
\includegraphics[width=8cm]{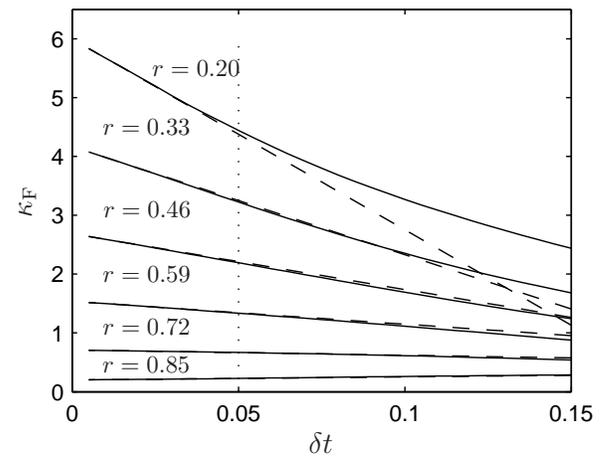}
\caption{
The force covariance $\kappa_{\text F}$ measured in DPD simulation plotted as a function of $\delta t$, for different values of $r$ (solid lines). The simulation setup is the same as in Fig.~\ref{fig:dissForce_richardson}. It is clearly visible that $\kappa_{\text F}$ has a linear region for small $\delta t$ (marked by dotted vertical line), as expected from Eq.~(\ref{eq:dpidpj_dt}). The dashed lines are linear functions with slopes given by the derivatives of $\kappa_{\text F}$ close to $\delta t = 0$.
}
\label{fig: Linear region}
\end{figure}

An example of the situation where we cannot measure $\kappa_{\text F}$ in the limit $\delta t \rightarrow0$ is shown in Fig.~\ref{fig:dissForce_richardson}. Here two measurements of $\kappa_{\text F}$ from a DPD simulation using time steps of different size ($\delta t_1=0.025$ and $\delta t_2=0.05$) deviates clearly from $\omega^2(r)$. The resulting estimate of $\omega^2(r)$,  obtained by Richardson extrapolation of $\kappa_{\text F}$ measurements, falls close on the original curve. It should be noted that $\kappa_{\text F}$ measurements are obtained from the same simulation (with time-step $\delta t=0.005$), as an increase in the DPD time step would alter the dynamics of the system. For a projected dynamics this is not a problem, as it evolves on the microscopic time scale. In Fig.~\ref{fig: Linear region} measurements of $\kappa_{\text F}$ from the same simulation is plotted against the size of the time difference between measurements, $\delta t$. As predicted by Eq.~(\ref{eq:dpidpj_dt}) the system exhibits a linear behaviour for small values of $\delta t$ (in this case $\delta t \lesssim 0.05$). Note, however, that how far the linear region extends varies significantly with the value of $r$.

\section{Coarse-graining of a Lennard-Jones fluid}
\label{sec:lj_example}

We now apply the method to a coarse-grained molecular system. 
Because of its simple form and because it is so well understood, we
examine the case of a coarse-grained two-dimensional Lennard-Jones (LJ) fluid. This is a single-species fluid, where the particles interact according to the standard pairwise potential
\begin{equation}\label{eq:LJ}
	V(r) = 4\epsilon \left[ (r/\sigma_\text{LJ})^{-12} -  (r/\sigma_\text{LJ})^{-6} \right].
\end{equation}
The parameters are chosen to correspond to bulk water at room pressure and temperature: the energy $\epsilon = 6.739$~meV, the interaction length $\sigma_\text{LJ} = 0.31655$~nm, and the mass $m_\text{LJ}=2.99\times 10^{-26}$~kg.

In DPD, the particles are often understood to be a collection of underlying particles, with properties such as mass and momentum defined from these. According to this view, we follow~\cite{flekkoy1}, where the coarse-grained dynamics is expressed in terms of a set of $N$ mesoscopic particles. Each particle has a position $\textbf{R}_k$, a velocity $\textbf{U}_k$, and a mass $M_k$. The instantaneous momentum of mesoscopic particle $k$ is defined as the sum of the momenta of the microscopic particles for which $k$ is the nearest mesoscopic particle, and the mass of the particle is defined as the total mass of these underlying particles:
\begin{eqnarray}
	\label{eq: projection}
	M_k &=& \displaystyle\sum_{i=1}^n\xi_k(\textbf{r}_i)m_i, \nonumber \\
	\textbf{P}_k &=& \displaystyle\sum_{i=1}^n \xi_k(\textbf{r}_i) m_i \textbf{v}_i, \\
	\textbf{U}_k &=&  \dot {\bf R}_k = \textbf{P}_k/M_k. \nonumber
\end{eqnarray} 
Here $n$ is the total number of microscopic particles, and $m_i$, $\textbf{r}_i$ and $\textbf{v}_i$ represent masses, positions and velocities, respectively, of the microscopic particles. $\xi_k(\textbf{r}_i)$ is $1$ if mesoscopic particle $k$ is closer to microscopic particle $i$ than any other mesoscopic particle is, and $0$ otherwise.

Though we use Eqs.~(\ref{eq: projection}) to find the motion of the mesoscopic particles, it is nevertheless illuminating to see how the effective forces acting on the mesoscopic particles are related to the forces acting on the microscopic particles.  Formally, we calculate the time-derivative of the momentum of the mesoscopic particle $k$ in Eq.~(\ref{eq: projection}). Between each passage of a microscopic particle from one mesoscopic particle to the next, $\xi_k({\bf r}_i)$ is constant (either zero or one). During these time intervals, the effective force acting on the mesoscopic particle is the sum of the forces acting on the microscopic particles closest to $k$:
\begin{equation} \label{eq:meso_force}
	M_k \dot {\bf U}_k = \sum_{i=1}^n \xi_k(\textbf{r}_i) {\bf f}_i ,
\end{equation}
Suppose microscopic particle $i$ leaves mesoscopic particle $k$. When this happens, the mesoscopic particle experiences an impulse
\begin{equation}
	{\bf I} = \Delta M_k {\bf U}_k = - m_i {\bf v}_i,
\end{equation}
so that the velocity of  mesoscopic particle $k$ changes instantaneously from ${\bf U}_k$ to
\begin{equation}
	{\bf U}'_k = \frac{1}{M_k - m_i} \left( M_k {\bf U}_k - m_i {\bf v}_i \right).
\end{equation}
The receiving mesoscopic particle is subject to the opposite impulse, $-{\bf I}$ (formally it is possible to express the force in terms of Dirac's delta function).

Finally, a word of caution: It might seem natural to use Eq.~(\ref{eq:meso_force}) alone to define the motion of the mesoscopic particles; however, in this dynamics the total momentum in the mesoscopic system changes when a microscopic particle moves from one mesoscopic particle to another.

\subsection{Estimated forces}

\begin{figure}
	\centering
	\includegraphics[width=8cm]{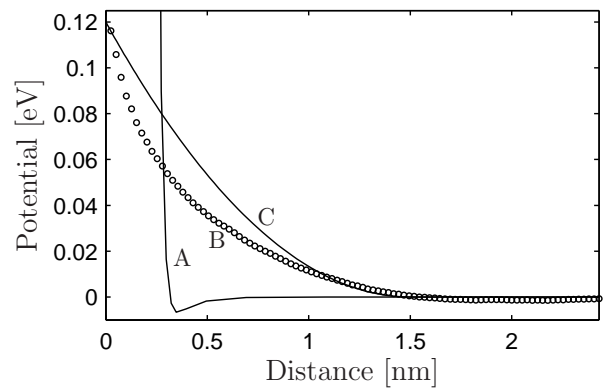}
\caption{\label{fig: Mesoscopic conservative potential}
A: Lennard-Jones potential used to simulate the microscopic particles. B: The effective potential for the coarse-grained system, obtained using the inverse MC method. C: The standard quadratic DPD potential, scaled to the same magnitude as the estimated potential. The main characteristics of DPD, soft-core repulsion and finite support, are confirmed by the retrieved potential.}
\end{figure}

The conservative interaction was determined from the RDF of the mesoscopic particles by the inverse MC method discussed earlier. The RDF was measured in LJ simulations with $1600$ particles in a simulation box with side length $12.48$~nm, temperature $333$~K and periodic boundaries. In the coarse-grained description, $160$ particles were used, resulting in an average of $10$ microscopic particles per mesoscopic particle. Fig. \ref{fig: Mesoscopic conservative potential} shows the potential compared with both the LJ potential (\ref{eq:LJ}) and the standard DPD potential (\ref{eq:linear_force}). The retrieved potential confirms the main characteristics of the standard DPD potential, i.e. soft-core repulsion and finite support.

\begin{figure}
	\centering
	\includegraphics[width=8cm]{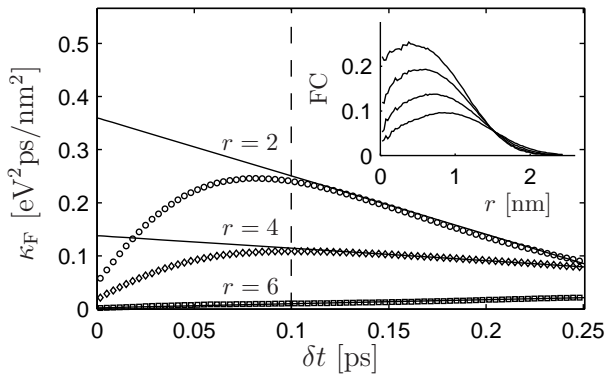}
\caption{\label{fig: dpidpjVsTime}
The force covariance $\kappa_{\text F}$ as a function of $\delta t$ (symbols), for three values of $r$ (shown by each curve). $\kappa_{\text F}$ is approximately linear for large enough $\delta t$ (to the right of the dashed line). For each value of $r$, the extrapolation of this region (indicated by solid lines) to $\delta t = 0$ determines the values of $\omega^2(r)$, c.f. Eq. (\ref{eq:dpidpj_dt}).
Inset: $\kappa_{\text F}$ as a function of $r$, for $\delta t = 0.1$ ps (top), $0.15$ ps, $0.2$ ps and $0.25$ ps (bottom).
It is clear from this figure that the terms proportional to $\delta t$ are not negligible in Eq.~(\ref{eq:dpidpj_dt}).
}
\end{figure}

\begin{figure}
	\centering
	\includegraphics[width=8cm]{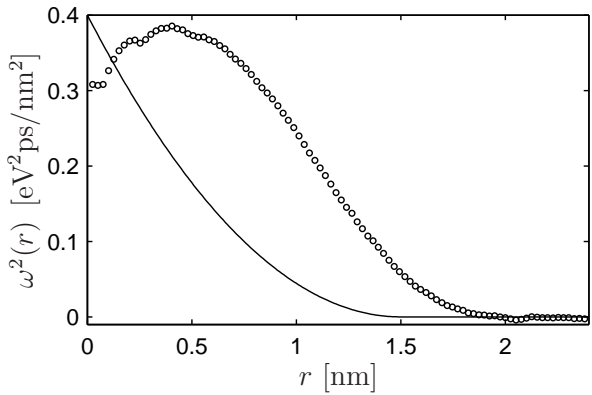}
\caption{\label{fig: Mesoscopic omega} 
The plotted circles show the estimate of $\omega^2(r)$, calculated from the force covariance $\kappa_{\text F}$ using Richardson extrapolation, c.f. Fig. \ref{fig: dpidpjVsTime}. The solid line is $\omega^2(r)$, using the DPD form in Eq.~(\ref{eq:linear_noise}), with the same cut off distance as for the conservative potential in Fig.~\ref{fig: Mesoscopic conservative potential}. The estimated stochastic interaction differs significantly from the function commonly used in DPD studies.}
\end{figure}

In Fig.~\ref{fig: dpidpjVsTime} we show $\kappa_{\text F}$ for the Lennard-Jones system as a function of $\delta t$ for different values of $r$ (the inset shows $\kappa_{\text F}$ as a function of $r$ for different values of $\delta t$). For small $\delta t$, $\kappa_{\text F}$ is increasing up to a maximum. We find a range starting at $\delta t = 0.1$~ps where $\kappa_{\text F}$ is approximately linear. In principle $\kappa_{\text F}$ is also linear for very small values of  $\delta t$, but since we know that the fluctuations we want to approximate with Markovian noise occurs on time scales $\lesssim 0.03$~ps (c.f section~\ref{sec:estimate_dissipative_term}), we reject this region. Hence, the linear region in the figure should match the linear region of Eq.~(\ref{eq:dpidpj_dt}). An extrapolation using data from the linear region gives the shape of $\omega^2(r)$ shown in Fig.~\ref{fig: Mesoscopic omega}. For comparison we have also included the standard shape of $\omega^2(r)$, which can be derived from Eq.~(\ref{eq:linear_noise}), with the magnitude scaled to fit the obtained $\omega^2(r)$ (solid line).  We note that the dissipative force derived from the mesoscopic particle motion is significantly broader than the standard shape. The conservative force is increasing only gradually as a pair of mesoscopic particles come within the interaction distance (approximately $1.5$~nm), the dissipative force grows much more rapidly.

\subsection{Consistency check}

\begin{figure}
	\centering
	\includegraphics[width=8.0cm]{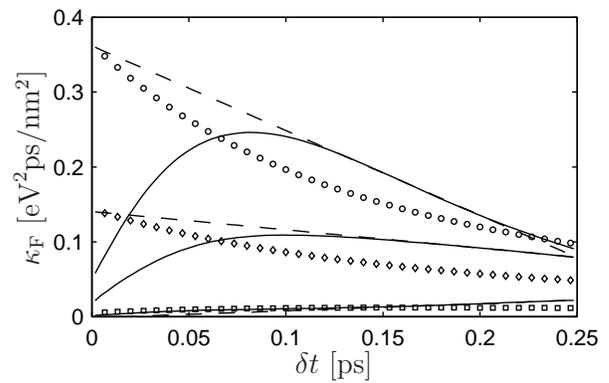}
\caption{\label{fig: dpd vs projection}
The plot shows measurements of the force covariance $\kappa_{\text F}$ plotted as a function of $\delta t$ for both the LJ-projection (solid lines) and a DPD simulation (symbols) using the estimated functional forms. Also plotted is the Richardson extrapolation of the LJ-projection (dashed lines). The curves are plotted for three values of $r$ (same as in Fig. \ref{fig: dpidpjVsTime}). The linear region of $\kappa_{\text F}$ for the DPD dynamics lies between  $\delta t = 0$ and $\delta t = 0.05$ ps, while the linear region of the projected dynamics is in the range $\delta t = 0.1$ to $\delta t = 0.25$  ps. $\kappa_{\text F}$ for the two systems do not have a coinciding linear region (for each value of $r$).}
\end{figure}

To test if the projected LJ-system can be represented by pairwise Langevinian dynamics, we perform a DPD simulation using the estimated functional forms of the conservative and dissipative forces, as shown in Fig.~\ref{fig: Mesoscopic conservative potential} and Fig.~\ref{fig: Mesoscopic omega}.  Setting up the DPD simulation so as to correspond to the projected LJ-dynamics, we obtained measurements of $\kappa_{\text F}$  for varying $\delta t$. In Fig.~\ref{fig: dpd vs projection}  the results (symbols) for a selection of $r$-values are plotted together with the corresponding measurements from the LJ-projection (solid lines), and the Richardson extrapolation of these (dashed lines). The linear region of $\kappa_{\text F}$ for the DPD dynamics lies between  $\delta t = 0$ and $\delta t \approx 0.05$ ps. As pointed out in section \ref{sec:estimate_dissipative_term}, this is in the region of the fast dynamics for the underlying system. Clearly, the linear regions for $\kappa_{\text F}$ in the DPD system and in the projected Lennard-Jones system do not coincide and therefore we cannot confirm that the projection can be formulated in terms of DPD.

As is seen in section~\ref{sec:estimate_dissipative_term}, the method we have developed works for any system that obeys the DPD ansatz. More strictly, it works for any system that evolves on a timescale where all terms of order $O(\delta t^2)$ can be neglected in  Eq. (\ref{eq:dpidpj_dt}). That allows $\omega(r)$ to be estimated either directly from $\kappa_{\text F}$ (if also first order terms of $\delta t$ are negligible), or through Richardson extrapolation of $\kappa_{\text F}$ for different values of $\delta t$. If we only have data available for the system on a longer timescale, there may not be a region where $\kappa_{\text F}$ is approximately linear. In this case it is not possible to use the linearization procedure to extract the dissipative force.

An important remark is that, in general, it cannot be concluded from measurements of $\kappa_{\text F}$ alone if the dynamics is Markovian or follows the DPD ansatz; a cross check with a DPD simulation is necessary. In the case of the projected LJ-dynamics, it was reasonable to assume that the linear region (see Fig.~\ref{fig: dpidpjVsTime}) could  be interpreted as the right timescale to consider for extracting the functional forms. However, $\kappa_{\text F}$ from the DPD simulation turned out to have its linear region on a  much shorter timescale than assumed for the projected dynamics.  

In the light of this result, there are two explanations for the observed deviations from the DPD ansatz, which differ with respect to whether the DPD ansatz is correct or not. If we first assume that the projected system follows the DPD ansatz, our guess of a linear region is not correct, and it follows that higher order terms of $\delta t$ will affect the value of $\kappa_{\text F}$, rendering our method unapplicable for this case. The solution to this problem calls for more sophisticated methods to estimate $\omega(r)$ from $\kappa_{\text F}$. The second possibility is that the projection does not produce a dynamics that follows the DPD ansatz. This could either simply be a result of our choice of projection, or it could point to deeper problems with constructing a coarse-graining scheme that leads to the DPD model. 

Flekk\o y et al.~\cite{flekkoy1, flekkoy2,flekkoy3} have used the same type of projection as presented in this article, c.f. Eq.~(\ref{eq: projection}), but rather than considering the coarse grained entities as spherical particles, they consider them as cells on a Voronoi lattice.
Within each cell, the fluid is assumed to correspond to an ideal fluid at a given pressure, temperature and velocity. Because of this, the system is similar in spirit to the Lattice-Boltzmann coarse-graining, but with dynamic cells. 
An advantage of this method is that the dissipative part of the evolution equations can be derived theoretically~\cite{flekkoy3}. However, this method involves keeping track of, and updating, the Voronoi lattice at each time step of simulation, rendering this technique much slower than standard DPD. As the construction of the Voronoi lattice depends explicitly on all particle positions in the simulation, it also introduces a need for higher order interactions than the simple pairwise central forces normally associated with DPD. If it proves impossible to find a projection giving rise to DPD dynamics on the coarse grained level (which is a question that calls for further investigation), this alternative approach might still provide a reasonable path to take for performing reliable mesoscopic simulations.

\section{Choice of projection}

\begin{figure}
\centering
\includegraphics[width=8cm]{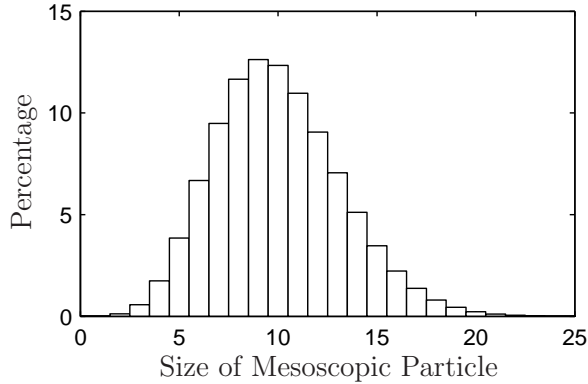}
\caption{\label{fig: mass distribution}Size distribution of mesoscopic particle masses, measured in units of microscopic particle masses. The data was obtained from a simulation using $1600$ microscopic and $160$ mesoscopic particles, giving an average mesoscopic particle size of $10$.}
\end{figure}

Although the projection used in this study, Eqs.~(\ref{eq: projection}), seems like a natural choice for DPD, it is a problem that the positions of the coarse-grained particles only weakly reflect the positions of the underlying particles. As is shown in Fig.~\ref{fig: mass distribution}, the number of microscopic particles per coarse-grained particle (i.e. the mass of the coarse-grained particle) exhibits large fluctuations, in sharp contrast to the standard DPD model where the masses of all particles are assumed to be equal and constant in time.

\begin{figure}
	\centering
	\includegraphics[width=8cm]{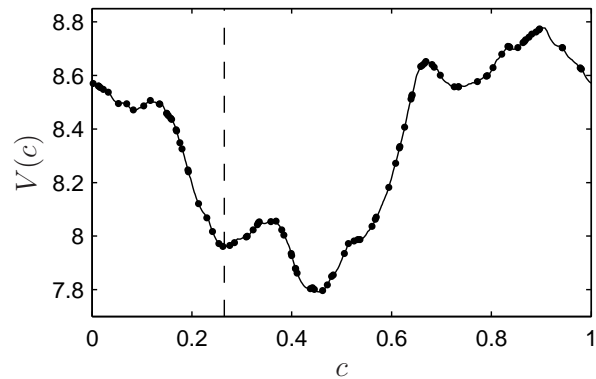}
\caption{\label{fig: sumOfSquares} The figure illustrates how the $k$-means clustering algorithm works. A minimum in the potential landscape corresponding to a local minimum in the sum of squares function, Eq.~(\ref{eq: sumOfSquares}) is found (indicated by vertical line). The small dots represent particle positions, $x_i$ in Eq.~(\ref{eq: sumOfSquares}), and $V(c)$ is the sum of squares distance as a function of cluster centre position. This example is made for illustrative purposes and therefore contains only a single cluster centre to which all the particles belong. As the particles move, $V(c)$ changes, with the effect that local minima are continuously created and destroyed. This process results in discontinuous trajectories for the cluster centre positions. }
\end{figure}

One way to make the coarse-grained particles more closely reflect the density variations in the underlying system of microscopic particles is to change the projection to incorporate movement of coarse-grained particles towards regions of higher particle concentrations. This can be achieved by using for instance the standard $k$-means clustering method~\cite{Classification} (or any other position based clustering algorithm) to calculate the positions of the coarse-grained particles given the positions of the underlying particles. This results in a model where the coarse-grained particles can be seen as clusters of underlying particles, with each cluster centre representing a local concentration peak of microscopic particles. An implementation has been made using this projection, and the results reveal some new difficulties not easily foreseen in advance. With this type of projection, the cluster centres move in a potential landscape of the kind depicted in Fig.~\ref{fig: sumOfSquares} for a one-dimensional system, where each local minimum of the curve represents a possible cluster centre position. The simulation was made for illustrative purposes, with a single cluster centre in a one-dimensional box with $N = 100$ particles, and with periodic boundary conditions. The curve represents the sum-of-squares distance from all the particles to the cluster centre, i.e.
\begin{equation}
\label{eq: sumOfSquares}
V = \displaystyle\sum_{i=1}^N {\min}\big(|c-x_i|^2, |L - c + x_i|^2\big),
\end{equation}
where the minimum of  the distance between the cluster and all periodically displaced images of particle $i$ is used. 

By differentiating Eq.~(\ref{eq: sumOfSquares}) with respect to the cluster centre position, $c$, it is easily shown that a minimum in the sum of squares function represents a local average of the positions of the microscopic particles belonging to the cluster. This information is just what the $k$-mean clustering algorithm uses to calculate the cluster centre positions. The fact that the example in Fig.~\ref{fig: sumOfSquares} contains only one cluster to which all the particles belong does not change the qualitative outcome that several local minima exist in the potential landscape. The result of this is inevitably that the cluster centre positions, represented by a given minimum in the potential landscape, will move with that local minimum until it disappears, which happens frequently in the course of the simulation, for instance by the merging of two originally separated minima. At this point, the cluster centre will jump to the adjacent minima, and in doing so it affects the neighboring cluster centres,  resulting in discontinuous movement of the coarse-grained particles. As discontinuous particle movement on the coarse-grained level, due only to strictly local interactions on the microscopic level, is highly unsatisfactory, this type of position based projections also leave much to be desired.

\section{Summary}

In this article we have developed a method for estimating the forces between particles in a system that evolves according to the DPD ansatz, i.e. Langevinian dynamics with pairwise central forces. The method works well for estimating both conservative and dissipative forces (with the stochastic force given by the dissipative through a fluctuation-dissipation theorem), and should work on any system that follows the DPD ansatz, as long as the time scale is small enough to let $\omega(r)$ be estimated from the force covariance $\kappa_{\text F}$. When applied to a projected dynamics of a Lennard-Jones system, we cannot conclude that the projection results in a DPD-like dynamics. The result points towards two possibilities: Either the projected dynamics is DPD-like, but outside the reach of our method, or in the worst case, there might be problems considering DPD as the result of a systematic coarse-graining method.

A natural extension of the work presented in this article is to examine systems where artifacts due to fluctuating mass and identity problems are not encountered, such as the frequently used united atoms approach. A simple example would be to coarse-grain water by letting the coarse-grained particle be the whole water molecule. Some work in this direction has already been made by \citet{Praprotnik2007b}. Another direction is to develop a more sophisticated method for estimating $\omega(r)$ from $\kappa_{\text F}$.

As we suspect that the linear ansatz for $\kappa_{\text F}$ is too simple, one might be tempted to simply use polynomials of higher degree in $\delta t$ and do a regression for the coefficent for each value of $r$ separately, based on the region where we think the DPD theory is valid. Since $\kappa_{\text F}$ is close to linear in this region, however, the result of extrapolating the resulting function to find the intersection with the $\delta t = 0$ axis may be rather sensitive to the precise choice of region in $\delta t$, and to noise in the measurement of $\kappa_{\text F}$ (from the finite number of samples).

Rather, one may consider going in the other direction: for a given choice of $\omega(r)$ (and keeping the conservative force fixed) we measure $\kappa_{\text F}$ as a function of $\delta t$ and $r$ and calculate a distance between $\kappa_{\text F}$ from the DPD simulation and $\kappa_{\text F}$ from the microscopic simulations. We may then use some optimization procedure that does not require explicit calculation of derivatives, e.g. the classic Downhill Simplex method or Monte Carlo methods, to obtain better estimates for $\omega(r)$ (see e.g. \cite{numerical_recipes} for a review of different suitable optimization methods).

\noindent {\bf Acknowledgments}: This work was funded (in part) by the EU integrated project FP6-IST-FET PACE, by EMBIO, a European Project in the EU FP6 NEST Initiative, and by the Research Councils of Norway and Sweden. We thank the European Center for Living Technology (ECLT) in Venice, Italy, for providing excellent conditions for a workshop in the fall of 2006 where large parts of the work were mapped out. We also wish to thank the anonymous referees for valuable comments and literature references. 

\bibliography{MesoscopicSimulation_PR}

\end{document}